\documentstyle[12pt,preprint,aps,tighten,epsfig,floats]{revtex}

\begin{document}

\newcommand{\subtext}[1]{\mbox{\footnotesize #1}}
\newcommand{\mbf}[1]{\mbox{\boldmath $#1$}}
\newcommand{\bra}[1]{\left\langle #1 \right|}
\newcommand{\brak}[2]{\left\langle #1 | #2 \right\rangle}
\newcommand{\lambar}{\mbf{\lambda}}
\newcommand{\xoo}{x_{0,0}}
\newcommand{\ket}[1]{\left| #1 \right\rangle}
\newcommand{\nsp}{\hspace{-.5ex}}
\newcommand{\mtrx}[9]{\renewcommand{\arraystretch}{1}\left[ \scriptsize{\begin{array}{cccc}
#2 & #3 & #4 & #5 \\ #6 & #7 & #8 & #9 \\ \end{array}}\right]_{#1}
\renewcommand{\arraystretch}{3} }
\newcommand{\mtrxs}[9]{\renewcommand{\arraystretch}{1}\left[ \scriptsize{\begin{array}{cccc}
#2 & #3 & #4 & #5 \\ #6 & #7 & #8 & #9 \\ \end{array}}\right]_{#1}^2
\renewcommand{\arraystretch}{3} }
\newcommand{\limit}[2]{\lim_{#1 \rightarrow #2}} 
\newcommand{\eqn}[1]{(~\nsp\ref{#1}~\nsp)}
\newcommand{\fig}[1]{Fig.~#1}
\newcommand{\figs}[1]{Figs.~#1}
\newcommand{\subp}{_{\scriptscriptstyle +}}
\newcommand{\subm}{_{\scriptscriptstyle -}}
\newcommand{\myAbstract}[1]{\bigskip\begin{center}
					\parbox[t]{5.5in}{#1}
			            \end{center}\medskip}
                        
\renewcommand{\thefootnote}{\arabic{footnote}} 
\renewcommand{\arraystretch}{3}

\begin{center}
{\huge A nonperturbative study of three-dimensional $\phi^4$ theory} 
\end{center}

\medskip
\begin{center}
{\large Mark Windoloski\footnote{\normalsize email: markw@het2.physics.umass.edu} \\
University of Massachusetts \\
Amherst, MA 01003}
\end{center}

\myAbstract{The spherical field formalism---a nonperturbative approach 
to quantum field theory---was recently introduced and applied to $\phi^4$
theory in two dimensions.  The spherical field method reduces a quantum
field theory to a finite-dimensional quantum mechanical system by 
expanding field configurations in terms of spherical partial wave modes.
We extend the formalism to $\phi^4$ theory in three dimensions and
demonstrate the application of the method by analyzing the phase 
structure of this theory.}

\section{Introduction}

Recently a new nonperturbative approach to quantum field theory, called spherical
field theory (SFT), has been introduced by Lee \cite{DLee}.  The idea of this
approach is to reduce a quantum field theory to a finite-dimensional quantum
mechanical system by expanding field configurations using spherical partial wave
modes.  The dimension of the system is made finite by truncating the partial wave
expansion by neglecting high spin waves.  The resulting quantum mechanical system
can then be analyzed numerically in Euclidean spacetime with the use of diffusion 
Monte Carlo methods (DMC).

SFT was developed in the context of $(\phi^4)_2$ theory  
in \cite{DLee}. The purpose of the present work is to demonstrate that SFT
can be extended beyond two-dimensional theories by developing the formalism
for $(\phi^4)_3$ theory.  In fact, we start out with a general extension 
of \cite{DLee} to an integer number of 
dimensions $N>2$ and specialize to $N=3$ when we renormalize the theory.

The paper is organized as follows:  We begin by expanding the field configurations
in the generating functional for $\phi^4$ theory in $N$ dimensions in terms of
spherical partial waves and integrating over the angular coordinates in the
resulting expression. Then, allowing the radial
coordinate to play the role of a time coordinate, we use the Feynman-Kac formula 
to write the generating functional in terms of a Schr\"{o}dinger time evolution 
operator called the spherical-field hamiltonian.
Next we apply functional differentiation to the generating functional to yield
the free two-point correlators of the partial wave fields.
Then we discuss the renormalization of 
$(\phi^4)_3$ in the context of spherical field theory.  Finally, to illustrate
the application of this new method we model the quantum mechanical system
defined by the spherical-field hamiltonian using DMC in order to analyze the
phase structure of $(\phi^4)_3$.

Consistent with Magruder's study of $(\phi^4)_3$ theory \cite{Mag} we find
that the theory undergoes a transition from a broken-symmetry phase to a
symmetric phase if the mass-squared parameter $\mu^2$ is negative.  In the
case of $\mu^2$ positive Magruder's analysis does not determine if spontaneous
symmetry breaking occurs in this theory.
Several authors have investigated this question using various 
nonperturbative methods; the results of their analyses are, however, in disagreement.  
Cea and Tedesco \cite{Cea} studied a generalization of the Gaussian effective potential 
in which they calculated the two-loop corrections to the Gaussian approximation.  They 
found no spontaneous symmetry breaking.  In contrast, Stancu \cite{Stancu} 
used an expansion of the effective potential based on the variational Gaussian 
effective potential method and found that a phase transition does exist.  Peter, 
H\"{a}user, Thoma and Cassing \cite{Peter} also found spontaneous symmetry breaking by 
applying the coupled set of equations of motion for Green functions up to 
the $4$-point level truncated by the use of the cluster expansions for $n$-point Green 
functions.  With our analysis we find that the theory
does not undergo a phase transition if $\mu^2$ is positive---the vacuum remains
invariant under the reflection $\phi\rightarrow -\phi$.

For the phase transition observed for $\mu^2<0$ we compute the critical 
coupling and the critical exponents $\nu$ and $\beta$.  Recent calculations of 
the critical exponents for the three-dimensional Ising universality class---to 
which $(\phi^4)_3$ is believed to belong---have been performed to a very high 
level of precision using various approaches such as Monte Carlo simulations 
of $\phi^4$ lattice models (MC) \cite{BFMS}, high temperature series 
expansions (HT) \cite{CPRV}, $d=3$ expansions and $\epsilon$-expansions \cite{GZJ}.
At the present early stage in the development of SFT we are limited by the 
underdeveloped state of the Monte Carlo algorithms available to model the
dynamics of systems governed by spherical-field hamiltonians
\footnote{This state of affairs is not restricted to SFT.  For example, see \cite{HBS} 
for a relevant discussion in the context of hamiltonian lattice gauge theory.}
and so we do not intend our estimates of $\nu$ and $\beta$ to be competitive with
the state-of-the-art results.  Rather, we compare our estimates with them
in an attempt to build confidence in this new method. 

There are important advantages to the spherical field formalism.  The method 
maintains exact rotational invariance and it can be generalized
to other geometries as was recently done with a study of quantum field theory
on the noncommutative plane \cite{CDP}.  Most importantly, it is a continuum
approach rather than a discrete approximation method and so it is free of certain
problems which arise in lattice calculations such as fermion doubling \cite{DLee2}.

\section{The Generating Functional}

The first step in the spherical field theory approach is to perform a spherical partial wave 
expansion of the fields in the generating functional.\footnote{It should be noted that 
covariant Euclidean quantization, an important
component of the spherical field approach, was introduced in \cite{FHJ}.} 
In $N$-dimensional Euclidean space the generating functional for $\phi^4$ theory 
is given by
\begin{equation}
Z[{\cal J}] = \int {\cal D}\phi \exp\left\{\int d^N\!\mbf{x}\, [{\cal L}+{\cal J}\phi]\right\}
\label{defZ}
\end{equation}
where
\begin{equation}
{\cal L} = -\frac{1}{2}\sum_{i=1}^{N}(\partial_i\phi)^2 - \frac{\mu^2}{2}\phi^2 - 
\frac{g}{4!}\phi^4.
\label{L}
\end{equation}
We express the cartesian coordinates in terms of hyperspherical coordinates,
\begin{eqnarray}
x_1 & = & t \sin\vartheta_1\sin\vartheta_2\cdots\sin\vartheta_{N-2}\cos\varphi \nonumber \\
x_2 & = & t \sin\vartheta_1\sin\vartheta_2\cdots\sin\vartheta_{N-2}\sin\varphi \nonumber \\
x_3 & = & t \sin\vartheta_1\sin\vartheta_2\cdots\cos\vartheta_{N-2} \\
\vdots & & \vdots \nonumber \\
x_{N-1} & = & t \sin\vartheta_1\cos\vartheta_2 \nonumber \\
x_N & = & t \cos\vartheta_1, \nonumber
\end{eqnarray}
and decompose the field $\phi$ and the source ${\cal J}$ into partial waves:
\begin{equation}
\phi(\mbf{x}) = \sum_{\lambar,m} \phi_{\lambar,m}(t) Y_{\lambar,m}(\Omega),
\label{PhiSer}
\end{equation}
\begin{equation}
{\cal J}(\mbf{x}) = \sum_{\lambar,m} {\cal J}_{\lambar,m}(t) Y_{\lambar,m}(\Omega).
\label{JSer}
\end{equation}
We have written $\lambar$ for the collection of indices 
$\lambda_1$, $\lambda_2$,$\cdots$, 
$\lambda_{N-2}$ and $\Omega$ for the angular variables $\vartheta_1, \vartheta_2,\cdots,
\vartheta_{N-2}, \varphi$.  Here, and throughout this paper, the sum over $\lambar$ 
runs over all integer values of the $\lambda_i$ such that
\begin{equation}
\lambda_1\geq \lambda_2\geq \lambda_3\geq\cdots\geq\lambda_{N-2}\geq 0 
\label{lamVals}
\end{equation}
and the sum over $m$ runs over all integer values satisfying
\begin{equation}
-\lambda_{N-2}\leq m \leq \lambda_{N-2}. 
\label{mVals}
\end{equation}
The $Y_{\lambar,m}$ are the $N$-dimensional hyperspherical harmonics\footnote{We use the phase 
convention $Y_{\lambar,m}^{\ast}(\Omega) = (-1)^m Y_{\lambar,-m}(\Omega)$.} which reduce to 
the usual spherical harmonics for $N=3$.  For a useful reference on hyperspherical harmonics as
well as the hyperspherical Bessel functions we will need below, see \cite{Avery}. 

To express the generating functional in terms of the partial waves we make a 
change of integration variables in the functional integral 
of Eq.~\eqn{defZ} from $\phi$ to the set $\{\phi_{\lambar,m}\}$.  The Jacobian of 
this transformation is 
\begin{equation}
\det \left[\frac{\delta\phi(\mbf{x}')}
{\delta\phi_{\lambar,m}(t)}\right] = \det\left[\delta(t-t') Y_{\lambar,m}(\Omega)\right],
\end{equation}
which is independent of the $\phi_{\lambar,m}(t)$.  Thus this change of integration 
variables simply introduces an overall constant which will be irrelevant for our purposes.

Now, having inserted Eqs.~\eqn{PhiSer} and \eqn{JSer} into Eq.~\eqn{defZ}, 
we can simplify the 
expression for $Z[{\cal J}]$ by using the orthogonality condition of the hyperspherical harmonics,
\begin{equation}
\int d\Omega\,\, Y^{\ast}_{\lambar,m}(\Omega) Y_{\lambar',m'}(\Omega) = 
\delta_{\lambar,\lambar'}\delta_{m,m'},
\end{equation}
to integrate over $\Omega$ in the action.  Before we can do this we must eliminate the 
angular derivatives of the $Y_{\lambar,m}$ which appear in the kinetic term of $\cal{L}$. 
This can be accomplished by integrating this kinetic term by parts to yield
\begin{equation}
-\frac{1}{2}\int d^N\!\mbf{x}\, \sum_{i=1}^{N}(\partial_i\phi)^2 = \frac{1}{2}\int 
d^N\!\mbf{x}\,\, \phi\triangle\phi
\end{equation}
where $\triangle$ denotes the generalized Laplacian operator which can be written in 
terms of the generalized angular momentum operator $\Lambda^2$ as
\begin{equation}
\triangle = \frac{1}{t^{N-1}}\frac{\partial}{\partial t} t^{N-1} \frac{\partial}{\partial t}
 - \frac{1}{t^2}\Lambda^2.
\end{equation}
Now the angular derivatives have been removed since the hyperspherical harmonics are 
eigenfunctions of $\Lambda^2$ satisfying the relation
\begin{equation}
\Lambda^2 Y_{\lambar,m}(\Omega) = \ell(\ell+N-2) Y_{\lambar,m}(\Omega)
\end{equation}
where, for ease of notation, we have written $\ell$ for $\lambda_1$, the first
component of $\lambar$. 

In performing the integration over $\Omega$ of the interaction term we encounter an
integral of the product of four hyperspherical harmonics.  Let us make the definition
\begin{equation}
\mtrx{N}{\lambar_1}{\lambar_2}{\lambar_3}{\lambar_4}{m_1}{m_2}{m_3}{m_4} 
\equiv \frac{2\pi^{N/2}}{\Gamma(N/2)} \int d\Omega\, \, Y_{\lambar_1,m_1}(\Omega) 
Y_{\lambar_2,m_2}(\Omega) Y_{\lambar_3,m_3}(\Omega) Y_{\lambar_4,m_4}(\Omega) 
\end{equation}
where the prefactor to the integral, which is the area of an $N$-dimensional 
unit sphere, is introduced for later convenience.  
This integral can be calculated by using the expansion
\begin{equation}
Y_{\lambar_1,m_1}(\Omega) Y_{\lambar_2,m_2}(\Omega) = \sqrt{\frac{\Gamma(N/2)}{2\pi^{N/2}}} 
\sum_{\lambar,m} C_N(\lambar_1,m_1;\lambar_2,m_2;\lambar,m) Y_{\lambar,m}(\Omega)
\end{equation}
where again a prefactor has been introduced for convenience.  The coefficients, 
$C_N(\lambar_1,m_1;\lambar_2,m_2;\lambar,m)$, have been worked out in \cite{WenAvery}.  
They vanish unless $m_1+m_2=m$ and so it follows that $\mtrx{N}{\lambar_1}{\lambar_2}
{\lambar_3}{\lambar_4}{m_1}{m_2}{m_3}{m_4}$ vanishes unless $m_1+m_2+m_3+m_4=0$.
In particular, for $N=3$ we have
\begin{eqnarray}
\mtrx{3}{\ell_1}{\ell_2}{\ell_3}{\ell_4}{m_1}{m_2}{m_3}{m_4} & 
\equiv & (-1)^{m_1+m_2+m_3}\delta_{m_4,-m_1-m_2-m_3}\sum_{\ell=|\ell_1-\ell_2|}^{\ell_1+\ell_2} 
C_3(\ell_1,m_1;\ell_2,m_2;\ell,m_1+m_2) \\
 & & \times C_3(\ell,m_1+m_2;\ell_3,m_3;\ell_4,m_1+m_2+m_3) \nonumber
\end{eqnarray}
with
\begin{eqnarray}
C_3(\ell_1,m_1;\ell_2,m_2;\ell,m) & = & \sqrt{\frac{(2\ell_1+1)(2\ell_2+1)}{(2\ell+1)}}
\langle \ell_1,\ell_2;m_1,m_2|\ell_1,\ell_2;\ell,m\rangle \nonumber \\ 
 & & \times \langle \ell_1,\ell_2;0,0|\ell_1,\ell_2;\ell,0\rangle. 
\end{eqnarray}

With these results we are in a position to write down our final expression for the 
generating functional.  We have
\begin{equation}
Z[{\cal J}] \propto \int \left(\prod_{\lambar,m} {\cal D}\phi_{\lambar,m}\right)
 \exp\left\{\int_{0}^{\infty} dt\, {\cal L}^{\subtext{Sph}}_{\cal J}\right\}
\label{J} 
\end{equation}
where ${\cal L}^{\subtext{Sph}}_{\cal J}$ is the spherical-field lagrangian,
\begin{eqnarray}
{\cal L}^{\subtext{Sph}}_{\cal J} & = & -\sum_{\lambar,m} (-1)^m \left( 
\frac{t^{N-1}}{2}\frac{d\phi_{\lambar,-m}}{dt}\frac{d\phi_{\lambar,m}}{dt} \right.
\label{LSph} \\
 & & +\left. \frac{\mu^2 t^2 + \ell(\ell+N-2)}{2} t^{N-3}\phi_{\lambar,-m}\phi_{\lambar,m} 
 - t^{N-1} {\cal J}_{\lambar,-m}\phi_{\lambar,m} \right) \nonumber \\ 
 & & -ct^{N-1}\sum_{\lambar_1,m_1}\sum_{\lambar_2,m_2}\sum_{\lambar_3,m_3}\sum_{\lambar_4}
\mtrx{N}{\lambar_1}{\lambar_2}{\lambar_3}{\lambar_4}{m_1}{m_2}{m_3}{-m_1\!-\!m_2\!-\!m_3} 
\nonumber \\
 & & \times\phi_{\lambar_1,m_1}\phi_{\lambar_2,m_2}\phi_{\lambar_3,m_3}
\phi_{\lambar_4,-m_1-m_2-m_3} \nonumber 
\end{eqnarray}
and the constant $c$ is defined as
\begin{equation}
c \equiv \frac{g}{4!}\frac{\Gamma(N/2)}{2\pi^{N/2}}.
\end{equation}

\section{The Spherical-Field Hamiltonian}

With the Feynman-Kac formula we can interpret the functional integral Eq.~\eqn{J} as a 
time evolution equation where the radial coordinate $t$ now plays the role of a time 
parameter.

The Feynman-Kac formula, generalized to include the class of time-dependent interactions 
relevant here, connects the path integral representation of Euclideanized quantum
mechanics to the Schr\"{o}dinger representation, 
\begin{equation}
\int_{\stackrel{\scriptstyle \phi(t_I)=x_I}{\phi(t_F)=x_F}} {\cal D}\phi\, \exp\left\{-
\int_{t_I}^{t_F} dt\, \left[\frac{a(t)}{2}\left(\frac{\partial\phi}{\partial t}\right)^2
 + V(\phi,t)\right]\right\} 
\propto \bra{x_F} T\, \exp\left\{-\int_{t_I}^{t_F} dt\, H(t)\right\}\ket{x_I} 
\end{equation}
where
\begin{equation}
H(t) = -\frac{1}{2a(t)}\frac{d^2}{dq^2} + V(q,t).
\end{equation}
We may use this formula to write down the Schr\"{o}dinger representation of Eq.~\eqn{J},  
\begin{equation}
Z[{\cal J}] \propto \bra{0}T\,\exp\left\{-\int_{0}^{\infty}dt\, H^{\subtext{Sph}}_{\cal J}(t)\right\}
\ket{a}
\label{JSchro}
\end{equation}
where 
\begin{eqnarray}
H^{\subtext{Sph}}_{\cal J}(t) & = & \sum_{\lambar,m} (-1)^m\left(-\frac{1}{2t^{N-1}}
\frac{\partial}{\partial q_{\lambar,-m}}\frac{\partial}{\partial q_{\lambar,m}} \right. 
\label{HJ} \\
 & & +\left. \frac{\mu^2 t^2 + \ell(\ell+N-2)}{2} t^{N-3} q_{\lambar,-m} q_{\lambar,m} - t^{N-1} 
{\cal J}_{\lambar,-m} q_{\lambar,m}\right) \nonumber \\
 & & +ct^{N-1}\!\!\sum_{\lambar_1,m_1}\sum_{\lambar_2,m_2}\sum_{\lambar_3,m_3}\sum_{\lambar_4}
\mtrx{N}{\lambar_1}{\lambar_2}{\lambar_3}{\lambar_4}{m_1}{m_2}{m_3}{-m_1\!-\!m_2\!-\!m_3}
\nonumber \\
 & & \times q_{\lambar_1,m_1} q_{\lambar_2,m_2} q_{\lambar_3,m_3} q_{\lambar_4,-m_1-m_2-m_3}
\nonumber
\end{eqnarray}
is a Schr\"{o}dinger time evolution generator which we will refer to as 
the spherical-field hamiltonian.
Following \cite{DLee} we take the quantum state at $t_I=0$ to be a superposition of
all possible states with each state equally weighted, 
\begin{equation}
\ket{a} \equiv \int_{-\infty}^{\infty}\left(\prod_{\lambar,m} dx_{\lambar,m}\right)
\ket{\{x_{\lambar',m'}\}}.
\end{equation}
In the set $\{ x_{\lambar',m'}\}$, the $\lambda_i'$ and $m'$ take all integer values
satisfying the inequalities \eqn{lamVals} and \eqn{mVals} respectively.
Note that, with this notation,  
\begin{equation}
q_{\lambar,m}\ket{\{x_{\lambar',m'}\}} = x_{\lambar,m}\ket{\{x_{\lambar',m'}\}}.
\end{equation}
The proper quantum state for $t_F\rightarrow\infty$
is the ground state, $\ket{0}$. 

We are faced with the difficulty that the spherical-field hamiltonian  
is written as an infinite series.  In practice we cut this series off 
by neglecting partial waves with spin higher than some value we call 
$J_{\subtext{max}}$.  Such an approximation is justified because, as pointed out in 
\cite{DLee}, the high spin partial waves correspond to high tangential momentum modes
so that a partial-wave cutoff $J_{\subtext{max}}$ corresponds to a momentum cutoff  
\begin{equation}
\Lambda^2 \sim \frac{J_{\subtext{max}}^2}{t^2}.
\end{equation}
Therefore the high spin modes will decouple from the theory, just as other 
high momentum effects, if 
the theory is renormalized to remove ultraviolet divergences.  Below, we will describe 
the proper method of renormalizing $(\phi^4)_3$ in the context of spherical field
theory, but to do so we need to derive an expression for the free two-point correlators 
of the partial wave fields which can then be used to evaluate the spherical-field version
of Feynman diagrams.

\section{Correlators}

Correlation functions for partial waves can be computed from the generating 
functional.  We begin by writing the generating functional Eq.~\eqn{defZ} as a power
series,
\begin{equation}
\frac{Z[{\cal J}]}{Z[0]} = 1+\frac{1}{2}\int\frac{d^N\!\mbf{p}\, d^N\!\mbf{x}\, d^N\!\mbf{y}}
{(2\pi)^N} e^{-i\mbf{p}\cdot (\mbf{x}-\mbf{y})} {\cal J}(\mbf{x}) {\cal J}(\mbf{y}) 
f(\mbf{p}^2) + \cdots
\label{ZSer}
\end{equation}
where $f(\mbf{p}^2)$ is the $\phi$ propagator,
\begin{equation}
f(\mbf{p}^{\, 2}) = \int d^N\!\mbf{x}\,\, e^{i\mbf{p}\cdot\mbf{x}}\bra{0}\phi (\mbf{x})
\phi(0)\ket{0}.
\end{equation}
We insert the partial wave expansion of ${\cal J}(\mbf{x})$ given in Eq.~\eqn{JSer} and 
expand the plane waves with
\begin{equation}
e^{-i\mbf{p}\cdot\mbf{x}} = (N-2)!! \frac{2\pi^{N/2}}{\Gamma(N/2)}\sum_{\lambar,m} (-i)^{\ell} 
j_{N,\ell}(kt) Y_{\lambar,m}^{\ast}(\Omega_k) Y_{n,m}(\Omega_t)
\label{PlnWv}
\end{equation}
where $k \equiv |\mbf{p}|$, $t \equiv |\mbf{x}|$ and $j_{N,n}$ is the
hyperspherical Bessel function of order $n$,
\begin{equation}
j_{N,n}(x) = \frac{\Gamma(\alpha)2^{\alpha-1}J_{n+\alpha}(x)}{(N-4)!!\,t^{\alpha}}, 
\end{equation}
where $\alpha\equiv\frac{N-2}{2}$.
Then we can perform the angular integration in Eq.~\eqn{ZSer} to obtain
\begin{eqnarray}
\frac{Z[{\cal J}]}{Z[0]} & = & 1+\frac{1}{2^{N-1}}\left[\frac{(N-2)!!}{\Gamma(N/2)}\right]^2
\sum_{\lambar,m} \int_{0}^{\infty} dk\, dt_x\, dt_y\, \\
& & \times k^{N-1} t_x^{N-1} t_y^{N-1} f(k^2) {\cal J}_{\lambar,-m}(t_x) 
{\cal J}_{\lambar,m}(t_y) j_{N,\ell}(k t_x) j_{N,\ell}(k t_y) + \cdots. \nonumber 
\end{eqnarray}
From this expression we may calculate the two-point correlator for $\phi_{\lambar,m}$,
\begin{eqnarray}
\bra{0}\phi_{\lambar,-m}(t_1)\phi_{\lambar,m}(t_2)\ket{0} \label{2ptNf} 
 & = & \frac{1}{Z[0]}\left.\left(\frac{\delta}
{t_1^{N-1}\delta {\cal J}_{\lambar,m}(t_1)}\frac{\delta}{t_2^{N-1}
\delta {\cal J}_{\lambar,-m}(t_2)} 
Z[{\cal J}]\right)\right|_{{\cal J}=0}  \\
 & = & \frac{1}{2^{N-2}}\left[\frac{(N-2)!!}{\Gamma(N/2)}\right]^2 (-1)^m \int_{0}^{\infty} 
dk\, k^{N-1} f(k^2) j_{N,\ell}(k t_1) j_{N,\ell}(k t_2). \nonumber
\end{eqnarray}

The free-field two-point $\phi_{\lambar,m}$ correlator is obtained by 
inserting the free propagator $f(k^2)=\frac{1}{k^2+\mu^2}$ into Eq.~\eqn{2ptNf}.
The result is  
\begin{eqnarray}
\bra{0}\phi_{\lambar,-m}(t_1)\phi_{\lambar,m}(t_2)\ket{0} & = & (-1)^m\mu^{2\alpha} 
\left[\frac{(N-2)!!\,\Gamma(\alpha)}{2\Gamma(N/2)}\right]^2 
\label{2ptNki} \\
 & & \times \left[ \theta(t_1-t_2) 
k_{N,\ell}(\mu t_1) i_{N,\ell}(\mu t_2) + \theta(t_2-t_1) k_{N,\ell}(\mu t_2) 
i_{N,\ell}(\mu t_1)\right], \nonumber
\end{eqnarray}
where $k_{N,n}$ and $j_{N,n}$ are the hyperspherical modified Bessel functions
of order $n$ which are given by
\begin{eqnarray}
k_{N,n}(x) & = & \frac{K_{n+\alpha}(x)}{\Gamma(\alpha)2^{\alpha-1} (N-4)!!\,t^{\alpha}}, \\
i_{N,n}(x) & = & \frac{\Gamma(\alpha)2^{\alpha-1}I_{n+\alpha}(x)}{(N-4)!!\,t^{\alpha}}.
\end{eqnarray}

\section{Renormalization}

Hereafter we will specialize to three dimensions.\footnote{We will no longer write 
dimension subscripts, {\it i.e.}\ $k_{3,n}(x)$ will be written as $k_n(x)$, etc.\ } 
In this case the 
theory can be made finite by adding a mass counterterm to the lagrangian Eq.~\eqn{L} 
which subtracts all primitively divergent self-energy diagrams.  In three dimensions 
there are two such diagrams.  One of these, shown in Fig.~\ref{fig:div}(a), is local and
so it is straightforward to write down the appropriate counterterms corresponding
to this diagram in the context of
spherical field theory.  We must explicitly subtract the contribution of all diagrams
having the form shown in Fig.~\ref{fig:sphdiv}(a).  
Using Eq.~\eqn{2ptNki}, and 
recalling the spherical-field lagrangian Eq.~\eqn{LSph}, we find that the counterterm 
which corresponds to a diagram of this form is proportional to  
\begin{equation}
(-1)^{m'} \mu ct^2 \mtrx{}{\ell}{\ell}{\ell'}{\ell'}{m}{-m}{m'}{-m'}
k_{\ell'}(\mu t)i_{\ell'}(\mu t) \phi_{\ell,-m}(t)\phi_{\ell,m}(t).
\label{SpherCT1}
\end{equation}
The proportionality constant is a product of two factors.  One is the degeneracy of  
the quartic term in the lagrangian which gives rise to such a diagram, or 
equivalently the number of distinct permutations of the columns in the bracket symbol.  
The other is the number of possible contractions of this quartic term which correspond 
to the diagram.  

The other divergent diagram is the so-called sunset diagram shown in 
Fig.~\ref{fig:div}(b).  This divergence can also be 
eliminated by using a local mass counterterm.  In ordinary field theory one finds that
the divergence is logarithmic, and so a single subtraction at any value of $k^2$ will
cancel it.  It is natural to choose $k^2=0$.    
In the context of  spherical field theory, we work in position space.  To
obtain a local counterterm we must consider all diagrams of the form shown in 
Fig.~\ref{fig:sphdiv}(b) 
and we must integrate over the coordinate of one of the vertices, say $t'$.  
However, we must do this in such a way as to keep our renormalization condition
translationally and rotationally invariant.

\input axodraw.sty
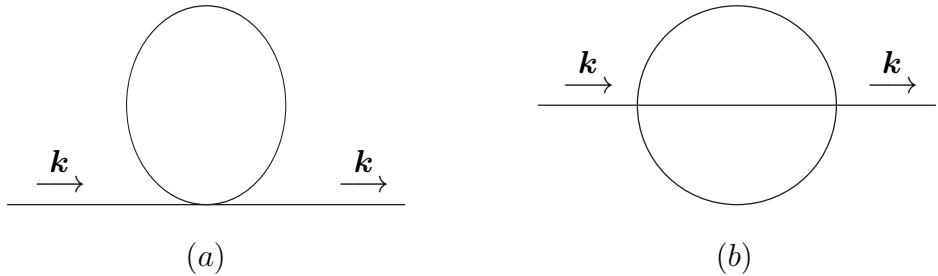
\begin{figure}
\vskip .1cm
\hskip 3.5cm

\begin{picture}(400,150)(-80,0)

\Line(-25,50)(125,50)
\Oval(50,87.5)(37.5,30)(0)
\put(-15,55){\shortstack{$\mbf{k}$\\$\longrightarrow$}}
\put(100,55){\shortstack{$\mbf{k}$\\$\longrightarrow$}}
\put(50,30){\makebox(0,0){$(a)$}}

\Line(175,87.5)(325,87.5)
\Oval(250,87.5)(37.5,37.5)(0)
\put(185,92.5){\shortstack{$\mbf{k}$\\$\longrightarrow$}}
\put(300,92.5){\shortstack{$\mbf{k}$\\$\longrightarrow$}}
\put(250,30){\makebox(0,0){$(b)$}}

\end{picture}
\caption{The primitively divergent self-energy diagrams
in ordinary perturbation theory.\hfill
\label{fig:div}}
\end{figure}
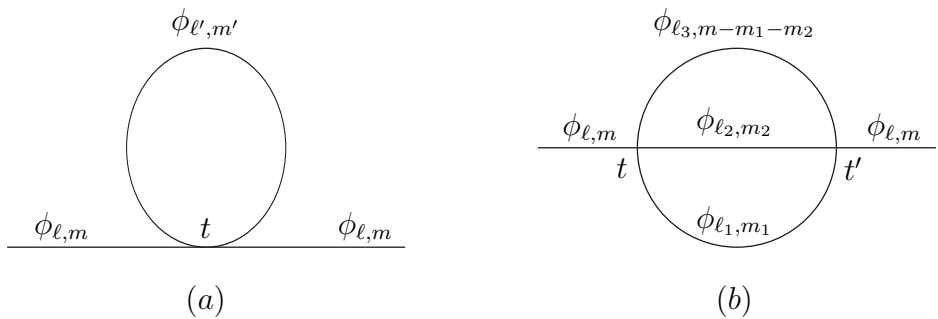
\begin{figure}
\vskip .1cm
\hskip 3.5cm

\begin{picture}(400,150)(-80,0)

\Line(-25,50)(125,50)
\Oval(50,87.5)(37.5,30)(0)
\put(-15,55){$\phi_{\ell,m}$}
\put(100,55){$\phi_{\ell,m}$}
\put(50,135){\makebox(0,0){$\phi_{\ell',m'}$}}
\put(50,57.5){\makebox(0,0){$t$}}
\put(50,30){\makebox(0,0){$(a)$}}

\Line(175,87.5)(325,87.5)
\Oval(250,87.5)(37.5,37.5)(0)
\put(185,92.5){$\phi_{\ell,m}$}
\put(300,92.5){$\phi_{\ell,m}$}
\put(250,135){\makebox(0,0){$\phi_{\ell_3,m-m_1-m_2}$}}
\put(250,97.5){\makebox(0,0){$\phi_{\ell_2,m_2}$}}
\put(250,60){\makebox(0,0){$\phi_{\ell_1,m_1}$}}
\put(207.5,80){\makebox(0,0){$t$}}
\put(295,80){\makebox(0,0){$t'$}}
\put(250,30){\makebox(0,0){$(b)$}}

\end{picture}
\caption{The primitively divergent self-energy diagrams
in spherical field theory.\hfill
\label{fig:sphdiv}}
\end{figure}

Working again in ordinary perturbation theory, we consider the sum of the sunset 
diagram and its counterterm in position space, 
\begin{equation}
\Sigma(\mbf{x},\mbf{x'}) = \sigma(\mbf{x},\mbf{x'}) - \delta\sigma(\mbf{x})\, 
\delta(\mbf{x}-\mbf{x'}).
\end{equation}
Here, the first term on the right hand side denotes the diagram itself and the 
second term denotes the
counterterm.  For convenience, we rewrite
the second term as $\delta\sigma(\mbf{x},\Omega_{\mbf{x'}})\, \frac{\delta(t-t')}
{t^2}$ absorbing the angular part of the delta function into $\delta\sigma$.  
We want the amplitude to vanish at $k^2=0$.  To impose this condition in a 
translationally and rotationally invariant way we demand that
\begin{equation}
\int d^3\mbf{x'}\, \Sigma(\mbf{x},\mbf{x'}) e^{-i\mbf{k}\cdot (\mbf{x}-\mbf{x'})} 
= f(k^2)
\end{equation}
where $f(k^2)$ is some function of $k^2$ which vanishes as $k^2\rightarrow 0$.

We can move the exponential factor independent of $\mbf{x'}$ to the right hand 
side of the equation and make use of Eq.~\eqn{PlnWv} to obtain the condition
\begin{equation}
\int d^3\mbf{x'}\, j_\ell(kt') Y_{\ell,m}(\Omega_{\mbf{x'}})\Sigma(\mbf{x},\mbf{x'}) 
= j_\ell(kt) Y_{\ell,m}(\Omega_{\mbf{x}}) f(k^2)
\end{equation}
which must hold for all possible values of $\ell$ and $m$.  Finally, we take the 
limit $k^2\rightarrow 0$.  In this limit, $j_\ell(kt)\propto (kt)^\ell$ and so we 
have
\begin{equation}
\int d\Omega_{\mbf{x'}}\, Y_{\ell,m}(\Omega_{\mbf{x'}}) \delta\sigma(\mbf{x},
\Omega_{\mbf{x'}}) = \frac{1}{t^\ell}\int_0^\infty dt'\, {t'}^{(\ell+2)}  
\int d\Omega_{\mbf{x'}}\, Y_{\ell,m}(\Omega_{\mbf{x'}}) \sigma(\mbf{x},\mbf{x'}) 
\end{equation}
for all $\ell$ and $m$.  This relation tells us the proper way to integrate 
a diagram of the form of Fig.~\ref{fig:sphdiv}(b) over $t'$ in order to obtain the 
corresponding 
counterterm.  With this information and again using Eq.~\eqn{2ptNki}, we can 
write down this counterterm.  It is proportional to 
\begin{equation}
-\frac{1}{2}(-1)^m c^2 t^2 \mtrxs{}{\ell_1}{\ell_2}{\ell_3}{\ell}{m_1}{m_2}
{m\!-\!m_1\!-\!m_2} {-m} I_{\ell_1,\ell_2,\ell_3}^{(\ell)}(\mu t)\phi_{\ell,-m}(t)
\phi_{\ell,m}(t)
\label{SpherCT2}
\end{equation}
where
\begin{eqnarray}
I_{\ell_1,\ell_2,\ell_3}^{(\ell)}(x) & = & k_{\ell_1}(x) k_{\ell_2}(x) 
k_{\ell_3}(x)\frac{1}{x^\ell}\int_{0}^{x}dx'\,{x'}^{(\ell+2)} i_{\ell_1}(x') 
i_{\ell_2}(x') i_{\ell_3}(x') \\
 & & +i_{\ell_1}(x) i_{\ell_2}(x) 
i_{\ell_3}(x)\frac{1}{x^\ell}\int_{x}^{\infty}dx'\,{x'}^{(\ell+2)} k_{\ell_1}(x') 
k_{\ell_2}(x') k_{\ell_3}(x'). \nonumber 
\end{eqnarray}
Again, the proportionality constant is a product of two factors.  Each diagram
of the form of Fig.~\ref{fig:sphdiv}(b) arises from the product of a quartic term 
from the 
lagrangian Eq.~\eqn{LSph} evaluated at coordinate $t$ and the same quartic term
evaluated at coordinate $t'$.  Therefore,  
one factor in the proportionality constant is the square of the degeneracy of    
this quartic term.  The other factor is the number of possible contractions of the
product of these quartic terms which correspond to the diagram.  

\section{The Phase Transition}

In \cite{Mag}, Magruder proved the existence of a phase transition in $(\phi^4)_3$ 
theory by extending Chang's proof \cite{Chang} for $(\phi^4)_2$ theory.  
Let us briefly recall Magruder's analysis.   
Consider the two lagrangians:
\begin{eqnarray}
{\cal L}\subp & = & \frac{1}{2}(\partial_\mu\phi)(\partial^\mu\phi)-\frac{1}{2}
\mu\subp^2\phi^2-4\pi c\phi^4+\frac{1}{2}\delta\mu\subp^2\phi^2, \\
{\cal L}\subm & = & \frac{1}{2}(\partial_\mu\phi)(\partial^\mu\phi)+\frac{1}{4}
\mu\subm^2\phi^2-4\pi c\phi^4+\frac{1}{2}\delta\mu\subm^2\phi^2 \nonumber
\end{eqnarray}
where $\mu\subp^2$ and $\mu\subm^2$ are both positive parameters.
The mass counterterms, $\delta\mu\subp^2$ and $\delta\mu\subm^2$, are chosen so that the  
theories are properly renormalized using the same conditions we have used above.  
Expressed in terms of a momentum cutoff $\Lambda$ they are defined as
\begin{eqnarray}
\delta\mu\subp^2 & = & 12 c (\Lambda-\mu\subp) + 96 c^2 \ln\left(\frac{\mu\subp}{\Lambda}
\right), \label{CTerms} \\  
\delta\mu\subm^2 & = & 12 c (\Lambda-\mu\subm) + 96 c^2 \ln\left(\frac{\mu\subm}{\Lambda}
\right). \nonumber
\end{eqnarray} 
The two lagrangians differ in the sign of their mass terms so that in the 
weak coupling limit the theory defined by ${\cal L}\subp$ has a vacuum with manifest 
symmetry under the reflection $\phi\rightarrow -\phi$ while the theory defined 
by ${\cal L}\subm$ has a vacuum with broken symmetry.
 
\begin{figure}
\vskip .1cm
\hskip 1.5cm
\psfig{figure=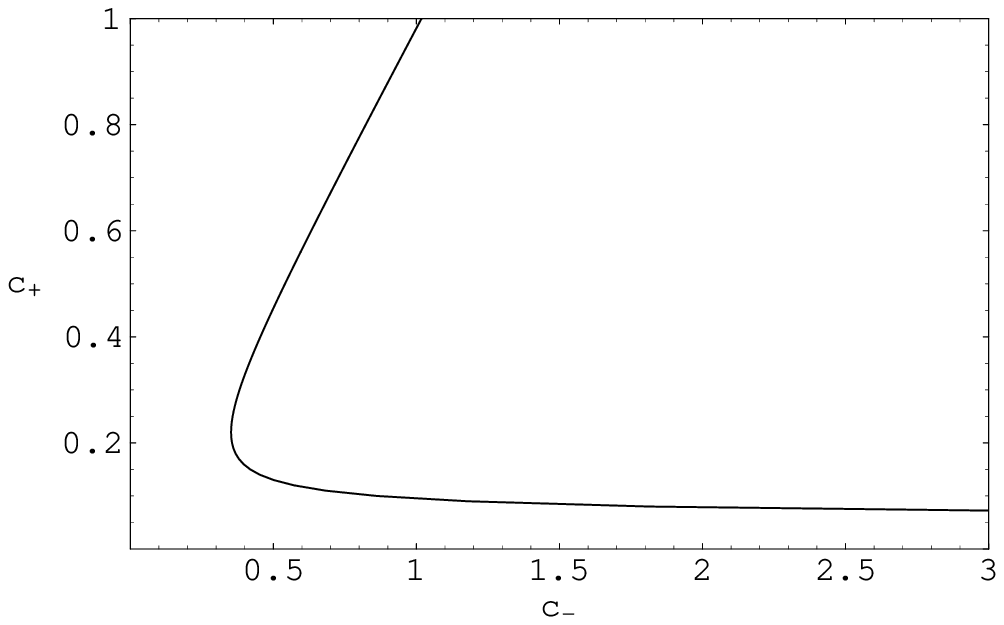,height=3.0in}
\caption{The functional relation between $c\subp\equiv\frac{c}{\mu\subp}$ and 
$c\subm\equiv\frac{c}{\mu\subm}$ as given by Eq.~\eqn{LIdent}.\hfill
\label{fig:magr}}
\end{figure}

The two lagrangians are identical if the parameters of the theories are 
chosen such that
\begin{equation}
-\mu\subp^2+\delta\mu\subp^2 = \frac{1}{2}\mu\subm^2+\delta\mu\subm^2.
\label{LIdent}
\end{equation}
The solution to this equation, shown graphically in Fig.~\ref{fig:magr},  
gives $c\subp\equiv\frac{c}{\mu\subp}$ as a function of $c\subm\equiv\frac{c}{\mu\subm}$.  
This solution has the property 
that where ${\cal L}\subp$ and ${\cal L}\subm$ overlap they are inversely 
related: a strong-coupling theory defined by ${\cal L}\subm$ is dual to a weak-coupling
theory defined by ${\cal L}\subp$.  
This implies that the theory defined by 
${\cal L}\subm$  undergoes a phase transition from a vacuum with broken symmetry to one 
with manifest symmetry at some finite value of the coupling strength, 
$c\subm=c\subm^{\subtext{crit}}$.  However, Fig.~\ref{fig:magr} shows that there is no 
overlap between
the two theories for $c\subm\lesssim 0.3522$.  Therefore, if 
$c\subm^{\subtext{crit}}$ is below this value the theory defined by 
${\cal L}\subp$ will have a vacuum with manifest symmetry for {\it any} value of  
$c\subp$.  On the other hand, if 
$c\subm^{\subtext{crit}}\gtrsim 0.3522$ then there is a finite range of 
intermediate coupling strengths for which the theory defined by ${\cal L}\subp$ will 
have a vacuum with broken symmetry.  Since this approach cannot tell us the 
value of $c\subm^{\subtext{crit}}$, it cannot determine whether or not spontaneous 
symmetry breaking exists in the theory defined by ${\cal L}\subp$.  

We can bring the method of spherical field theory to bear on this question by 
studying the quantum mechanical system defined by the spherical-field 
hamiltonian.  The appropriate hamiltonian is found by adding the two counterterms
discussed above to the spherical-field hamiltonian given by Eq.~\eqn{HJ} and 
setting ${\cal J} = 0$.  We can do this for 
the Euclideanized theory corresponding to ${\cal L}\subm$ as easily as for the one 
corresponding to ${\cal L}\subp$ since the renormalization procedure is the same for 
both theories as indicated by Eq.~\eqn{CTerms}.  The same counterterms, 
Eqs.~\eqn{SpherCT1} and 
\eqn{SpherCT2}, apply in both cases.  To avoid the complication of working with a
time-dependent hamiltonian we have chosen to investigate the original three-dimensional
system on a spherical surface of radius $R$.  Then we may set $t=R$ in the 
spherical-field hamiltonian, rendering it time-independent.  The full field theory result
is recovered in the limit $J_{\subtext{max}}\rightarrow\infty$, $R\rightarrow\infty$.

As an example, we display the spherical-field hamiltonian corresponding to 
${\cal L}\subp$ for $J_{\subtext{max}}=1$:
\begin{eqnarray}
H & = & -\frac{1}{2 R^2} \left( \frac{\partial^2}{\partial q_{0,0}^2} + 
\frac{\partial^2}{\partial q_{1,0}^2} + \frac{\partial^2}{\partial q_{1,s1}^2} + 
\frac{\partial^2}{\partial q_{1,a1}^2} \right) 
 + \frac{\mu^2 R^2}{2} q_{0,0}^2 + \frac{\mu^2 R^2 + 2}{2} \left( q_{1,0}^2
+ q_{1,s1}^2 + q_{1,a1}^2 \right) \nonumber \\
 & & +c R^2 \left[ q_{0,0}^4 + 6 q_{0,0}^2 \left( q_{1,0}^2 + q_{1,s1}^2
+ q_{1,a1}^2 \right) + \frac{\scriptstyle 9}{\scriptstyle 5} \left( 
q_{1,0}^2 + q_{1,s1}^2 + q_{1,a1}^2 \right)^2 \right] \\
 & & -6\mu c R^2 \left[ k_0(\mu R) i_0(\mu R) + 3 k_1(\mu R) i_1(\mu R) \right]
\left( q_{1,0}^2 + q_{1,s1}^2 + q_{1,a1}^2 \right) \nonumber \\
 & & +48 c^2 R^2 \left[ \left( I_{0,0,0}^{(0)}(\mu R) + 9 I_{0,1,1}^{(0)}(\mu R)
\right) q_{0,0}^2 \right. \nonumber \\ 
 & & \left. +3 \left( I_{0,0,1}^{(1)}(\mu R) + \frac{\scriptstyle 9}{\scriptstyle 5}
I_{1,1,1}^{(1)}(\mu R) \right)  \left( q_{1,0}^2 + q_{1,s1}^2 + q_{1,a1}^2 \right) 
\right] \nonumber
\end{eqnarray}
where
\begin{equation}
q_{1,s1}=\frac{q_{1,1}+q_{1,-1}}{i \sqrt{2}} \hspace{2em}\mbox{and}\hspace{2em}
 q_{1,a1}=\frac{q_{1,1}-q_{1,-1}}{\sqrt{2}}.
\end{equation}
The spherical-field hamiltonian corresponding to 
${\cal L}\subm$ for $J_{\subtext{max}}=1$ can be obtained from the above
expression by letting $\mu^2\rightarrow -\frac{1}{2}\mu^2$ in the mass terms while
leaving the counterterms unchanged.
Due to their lengths we will not include the corresponding expressions for higher
values of $J_{\subtext{max}}$ but they can be written down in a straightforward
way using Eqs.~\eqn{HJ}, \eqn{SpherCT1} and \eqn{SpherCT2}.

Equipped with the spherical-field hamiltonian we can use the method introduced 
in \cite{MRL} to find the vacuum expectation value and the physical mass as a
function of the coupling, $c$.  Near the phase transition the behavior of these
quantities define the critical exponents $\beta$ and $\nu$ respectively. 

We calculate the physical mass 
in the symmetric phase of the theory 
by using DMC\footnote{A self-contained 
introductory presentation of DMC is given in \cite{KFS}.}  
to compute the matrix element 
\begin{equation}
f(t)=\bra{b}q_{0,0} e^{-tH} q_{0,0}\ket{b},
\end{equation}
where $\ket{b}$ is any state that is even under the reflection transformation
$\phi\rightarrow-\phi$.  To see that one can extract the physical mass from 
$f(t)$ insert a complete set of energy eigenstates $\ket{i}$ corresponding to
energy eigenvalues $E_i$:
\begin{equation}
f(t)=\sum_i e^{-tE_i}\left| \bra{i}q_{0,0}\ket{b}\right|^2.
\end{equation}
Now the contribution of the ground state vanishes in this sum because both 
$\ket{0}$ and $\ket{b}$ are even under reflection symmetry while $q_{0,0}$
is odd.  Therefore, the sum is dominated by the one-particle-at-rest state
in the limit $t\rightarrow\infty$.  If we shift the energy scale such that
$E_0=0$ then for large $t$
\begin{equation}
f(t)\sim e^{-mt},
\end{equation}
where $m$ is the physical mass.

Next let us consider the vacuum expectation value.  
In the broken-symmetry
phase of the theory there are two degenerate ground states if the size of the
system becomes infinite, {\it i.e.}\ $R\rightarrow\infty$.  Let us take these two
ground states to be $\ket{0^+}$, which is non-zero only for $q_{0,0}>0$, and
$\ket{0^-}$, which is non-zero only for $q_{0,0}<0$.  We choose these such that
they are normalized to unity and so that under reflection symmetry
they transform from one to the other.  Now our task is to calculate the
vacuum expectation value $\bra{0^+}\phi\ket{0^+}$.  Using translational and rotational
invariance we can write this as
\begin{equation}
\bra{0^+}\phi\ket{0^+} = \frac{1}{4\pi} \int d\Omega\, \bra{0^+}
\phi(t,\theta,\varphi)\ket{0^+} = \bra{0^+}\frac{q_{0,0}}{\sqrt{4\pi}}
\ket{0^+}.
\end{equation} 
Let us rewrite the operator $q_{0,0}$ in terms of a 
projection operator,
\begin{equation}
q_{0,0} = \int^{\infty}_{-\infty} d\xoo\, \xoo\ket{\xoo}\bra{\xoo}.
\end{equation}
Now we have
\begin{equation}
\bra{0^+}\phi\ket{0^+} = \int_0^{\infty}d\xoo \, \xoo\frac{\left|\brak{\xoo}{0^+}
\right|^2}{\sqrt{4\pi}}
\end{equation}
where the lower limit of integration can be taken to be $\xoo=0$ since $\ket{0^+}$ 
vanishes for $\xoo <0$.  

The task which remains is to calculate the matrix element
which appears in this last expression.  To do this we need to know the ground state 
$\ket{0^+}$.  The time evolution operator $\exp{(-Ht)}$ 
operating on any state $\ket{b}$ will project out the state
of lowest energy as $t\rightarrow\infty$ 
provided $\ket{b}$ has a non-zero overlap with this lowest-energy state.  
Since we are working with a system of finite size the vacuum will not 
be exactly degenerate.  For finite $R$ the state that is lowest in energy is 
the symmetric superposition of $\ket{0^+}$ and $\ket{0^-}$,
\begin{equation}
\ket{0^s} = \frac{1}{\sqrt{2}}\left(\ket{0^+}+\ket{0^-}\right).
\end{equation}
Therefore we have
\begin{equation}
\left|\brak{\xoo}{0^s}\right|^2=\limit{t}{\infty}\frac{\bra{b}e^{-tH}
\ket{\xoo}\bra{\xoo}e^{-tH}\ket{b}}{\bra{b}e^{-2tH}\ket{b}},
\label{gofz}
\end{equation}
where $\ket{b}$ is chosen such that $\bra{0^s}b\rangle$ is not zero.
Now, since $\ket{0^+}=\sqrt{2}\ket{0^s}$ for $\xoo >0$, the vacuum expectation 
value can be written as
\begin{equation}
\bra{0^+}\phi\ket{0^+}=2\int_0^{\infty}d\xoo\,\, \xoo\,\, g(\xoo),
\end{equation}
where
\begin{equation}
g(\xoo)=\frac{\left|\brak{\xoo}{0^s}\right|^2}{\sqrt{4\pi}}.
\label{gDef}
\end{equation}
We can calculate $g(\xoo)$ by using DMC to evaluate 
the right-hand-side of Eq.~\eqn{gofz}.  However, we will not calculate $\bra{0^+}
\phi\ket{0^+}$ by integrating $2\xoo g(\xoo)$.  Instead we will take
\begin{equation}
\bra{0^+}\phi\ket{0^+}=\xoo^{\subtext{max}}
\end{equation}
where $\xoo^{\subtext{max}}$ is the non-negative value of $\xoo$ for which $g(\xoo)$ has
its maximum value.  This approach is justified because $g(\xoo)$ becomes sharply 
peaked as the size of the system becomes large and
$g(\xoo)$ satisfies 
\begin{equation}
\int_{-\infty}^{\infty}d\xoo\, g(\xoo)=2\int_0^{\infty}d\xoo\, g(\xoo)=1.
\end{equation} 
This approach converges to the same result as integrating $2\xoo g(\xoo)$ in the limit
$R\rightarrow\infty$, however, as indicated in \cite{MRL} it is less susceptible
to systematic errors for finite values of $R$.

\section{Results}

\begin{figure}
\vskip .1cm
\hskip 2.5cm
\psfig{figure=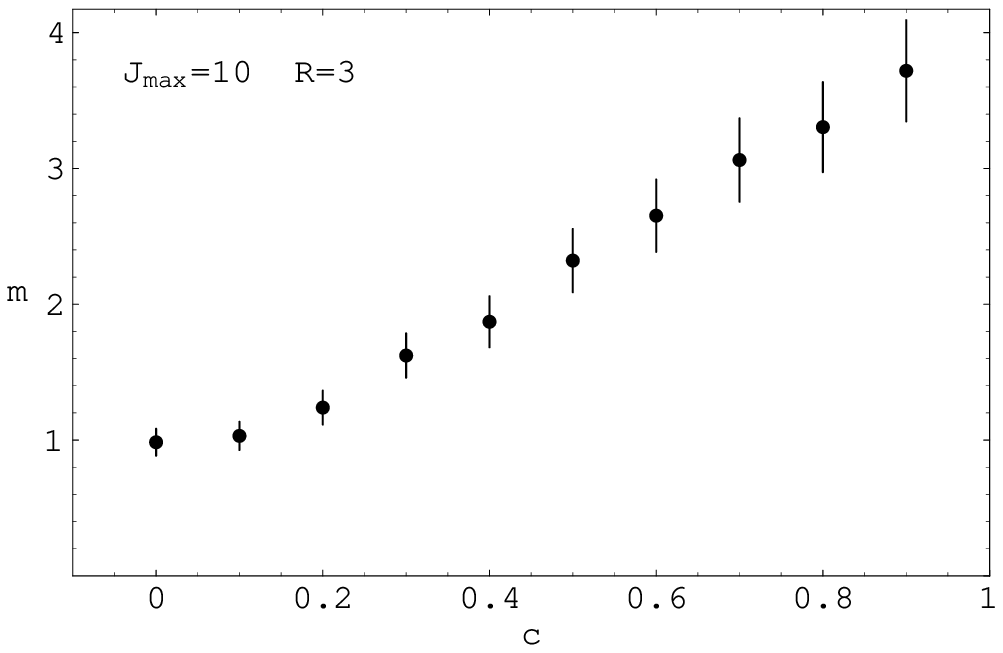,height=3.0in}
\caption{$m$ as a function of $c$ for the theory defined
by $\cal{L}\subp$ with $J_{\subtext{max}}=10$, $R=3$.\hfill
\label{fig:m103p}}
\end{figure}

\begin{figure}
\vskip .1cm
\hskip 1.2cm
\psfig{figure=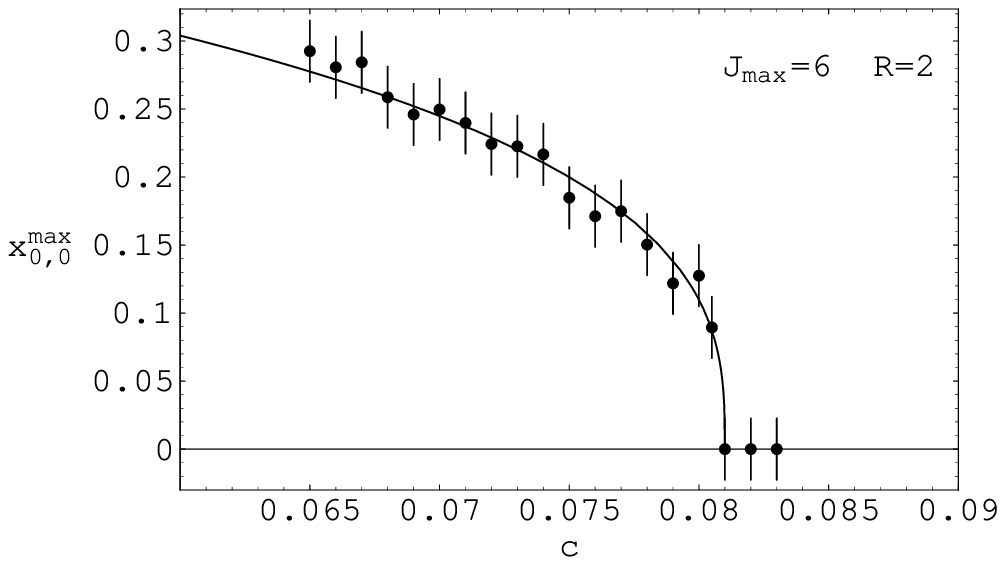,height=3.0in}
\caption{$x_{0,0}^{\subtext{max}}$ as a function of $c$ for the theory defined
by $\cal{L}\subm$ with $J_{\subtext{max}}=6$, $R=2$.\hfill
\label{fig:vev62}}
\end{figure}

\begin{figure}
\vskip .1cm
\hskip 1.2cm
\psfig{figure=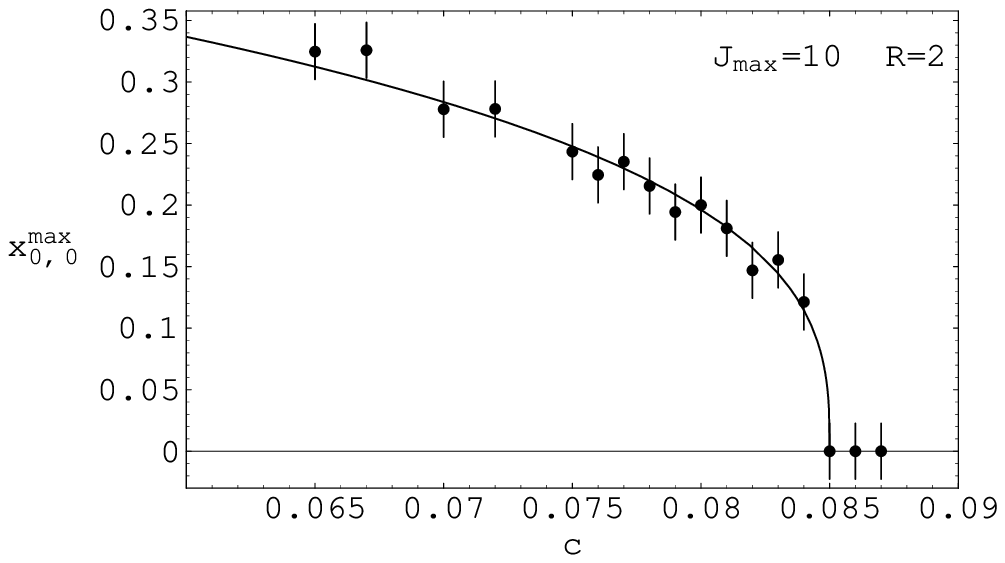,height=3.0in}
\caption{$x_{0,0}^{\subtext{max}}$ as a function of $c$ for the theory defined
by $\cal{L}\subm$ with $J_{\subtext{max}}=10$, $R=2$.\hfill
\label{fig:vev102}}
\end{figure}

\begin{figure}
\vskip .1cm
\hskip 1.2cm
\psfig{figure=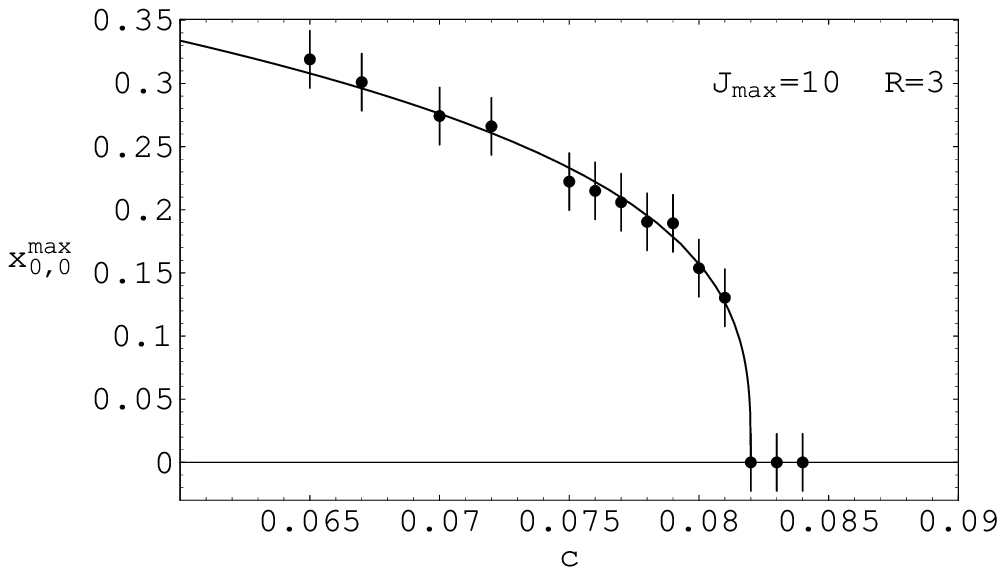,height=3.0in}
\caption{$x_{0,0}^{\subtext{max}}$ as a function of $c$ for the theory defined
by $\cal{L}\subm$ with $J_{\subtext{max}}=10$, $R=3$.\hfill
\label{fig:vev103}}
\end{figure}

\begin{figure}
\vskip .1cm
\hskip 1.2cm
\psfig{figure=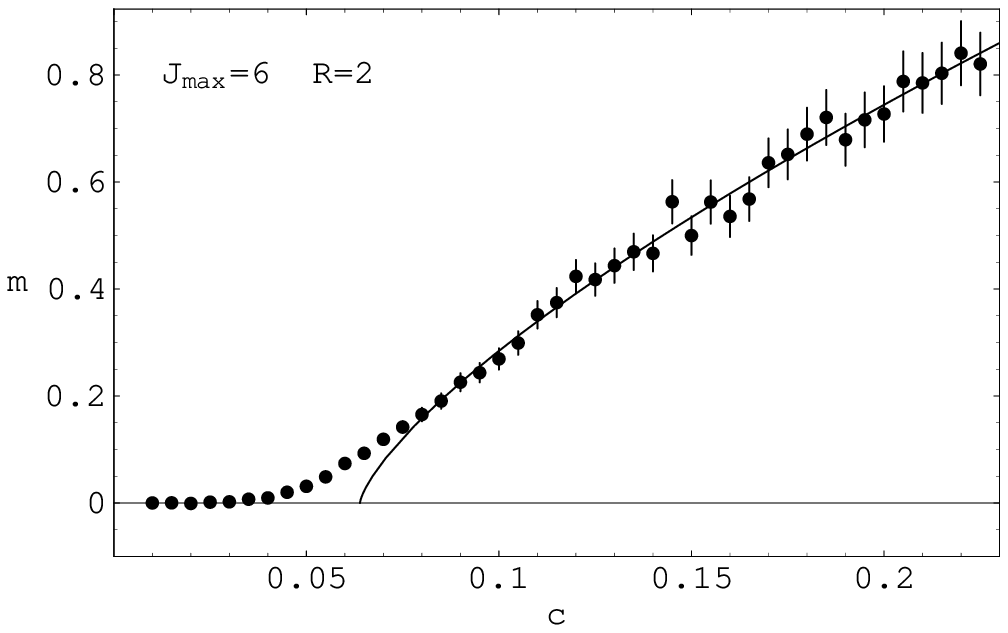,height=3.0in}
\caption{$m$ as a function of $c$ for the theory defined
by $\cal{L}\subm$ with $J_{\subtext{max}}=6$, $R=2$.\hfill
\label{fig:m62}}
\end{figure}

\begin{figure}
\vskip .1cm
\hskip 1.2cm
\psfig{figure=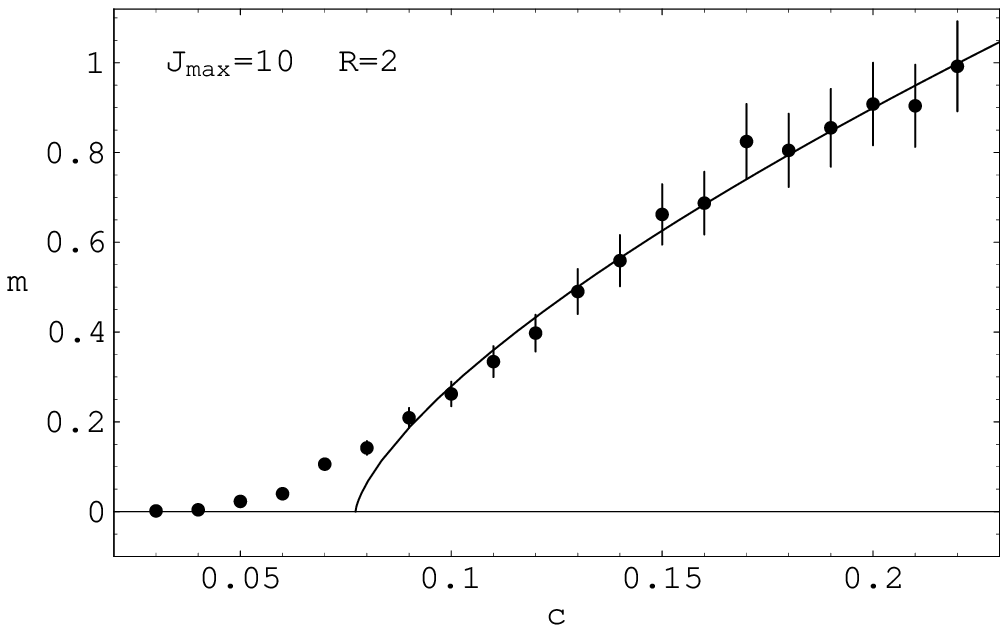,height=3.0in}
\caption{$m$ as a function of $c$ for the theory defined
by $\cal{L}\subm$ with $J_{\subtext{max}}=10$, $R=2$.\hfill
\label{fig:m102}}
\end{figure}

\begin{figure}
\vskip .1cm
\hskip 1.2cm
\psfig{figure=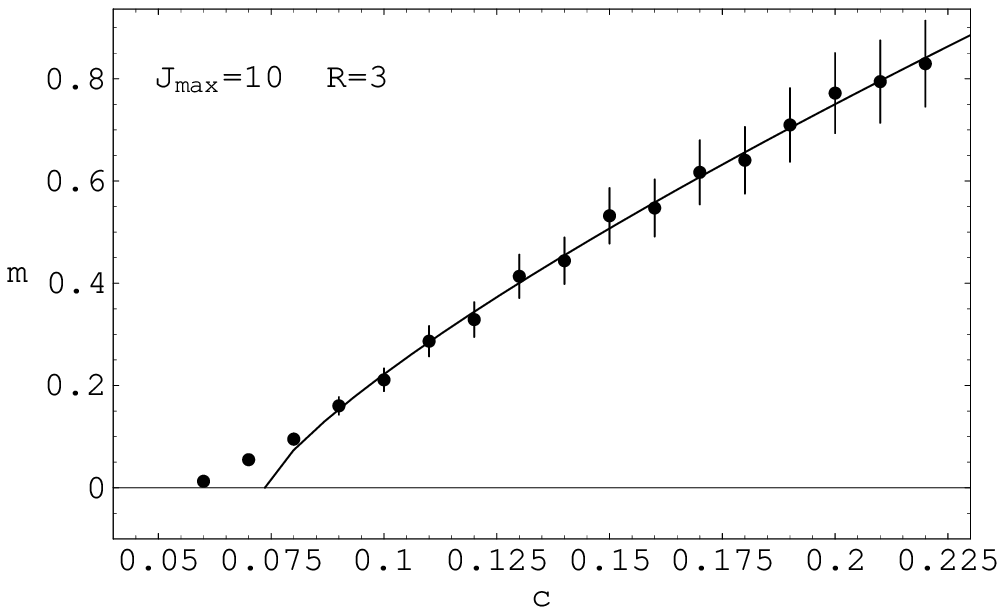,height=3.0in}
\caption{$m$ as a function of $c$ for the theory defined
by $\cal{L}\subm$ with $J_{\subtext{max}}=10$, $R=3$.\hfill
\label{fig:m103}}
\end{figure}

The results of our DMC calculations are presented in 
Figs.~\ref{fig:m103p}-\ref{fig:m103}.  The error bars indicate our best 
estimate of the error due to
statistical fluctuations as well as the contamination of higher energy states in
the physical mass calculations.

For the theory defined by $\cal{L}\subp$ 
we have calculated the vacuum expectation value and the physical mass over a range
of coupling strengths\footnote{We work in units such that $\mu=1$.}
 from weak coupling ($c\sim 0$) to very strong coupling ($c\sim 1$).  
This calculation was done with $J_{\subtext{max}}=10$ and $R=3$
which corresponds to $(J_{\subtext{max}}+1)^2=121$ partial wave modes and a 
spherical surface of area $4\pi R^2=36\pi$.
The vacuum expectation value is found to remain zero,  
{\it i.e.}\ the function $g(x_{0,0})$ has a single maximum at $x_{0,0}=0$, throughout
this range.  Furthermore, as shown in Fig.~\ref{fig:m103p}, the physical mass begins 
near unity at
weak coupling and consistently grows larger as the coupling strength is increased
to large values.  
We would expect the physical mass to vanish at the point where
$\phi\rightarrow -\phi$ reflection symmetry is spontaneously broken.  Our
results clearly show that there is no phase transition in the theory
corresponding to $\cal{L}\subp$.


On the other hand, our results do confirm the phase transition expected for the
theory defined by $\cal{L}\subm$.  We calculated the vacuum expectation value and
the physical mass using $J_{\subtext{max}}=6$, $R=2$; $J_{\subtext{max}}=10$, $R=2$;
and $J_{\subtext{max}}=10$, $R=3$.  The results are shown in Figs.~\ref{fig:vev62}
\nsp-\ref{fig:m103}.  
We find
that at strong coupling the vacuum expectation value is zero but as we decrease
the coupling strength the
theory develops a nonzero vacuum expectation value and the physical mass vanishes.  
This clearly indicates that
$\phi\rightarrow -\phi$ reflection symmetry, while manifest at strong coupling, is
spontaneously broken at weak coupling.

The critical exponent $\beta$ is defined by the behavior of the vacuum 
expectation value
as we approach the phase transition from the broken-symmetry phase
while the critical exponent $\nu$ is defined by the behavior of the physical mass $m$ 
as we approach the phase transition from the symmetric phase.
Therefore, we have fit the results using the parametrized forms
\begin{equation}
\bra{0^+}\phi\ket{0^+} = a (c_{\subtext{crit}}-c)^{\beta}
\hspace{0.25in} \mbox{and} \hspace{0.25in}
m = b (c-c_{\subtext{crit}})^{\nu}. 
\label{NuBetaDef}
\end{equation}
The results of these fits are shown in Tables~\ref{tabone} and \ref{tabtwo}.

The vacuum expectation value results are consistent with each other and with 
the recent calculations of $\beta$ shown in Table~\ref{tabthree}.
Also note that $c_{\subtext{crit}} <0.3522$.  
This is consistent with the finding of no phase transition for the $\cal{L}\subp$ 
theory as discussed below Eq.~\eqn{LIdent}.

The physical mass results are not as satisfying.  We obtain values for $\nu$ which
are consistently higher than expected for the three-dimensional Ising universality
class as seen by a comparison with the recent calculations of $\nu$ shown in
Table~\ref{tabthree}.
Furthermore, $c_{\subtext{crit}}$ is found to be lower than the value obtained from 
the vacuum expectation value calculations.  These difficulties arise because of a
systematic error in the calculation of the physical mass very near the phase transition
due to finite-size effects.  This systematic error is apparent in each of 
Figs.~\ref{fig:m62}-\ref{fig:m103} where the calculated values form a tail 
at small $c$.
In the region of these tails the physical mass corresponds to a correlation
length which is larger than the size of the system as given by $R$ and so these
points are of dubious validity.  Therefore we must do our fits only for those points 
which are not too near the phase transition, {\it i.e.}\ for which $m\gtrsim\frac{1}{\pi R}$.
This restriction makes it problematic to extract an accurate value for $\nu$ from
fits to the results.  
In the case of the vacuum expectation value calculations 
these finite-size effects are minimized as discussed below Eq.~\eqn{gDef}.

\renewcommand{\arraystretch}{1}
\begin{table}[tbh]
\begin{center}
\begin{tabular}{ccccc}

 $J_{\subtext{max}}$ & $R$ & $a$ & $c_{\subtext{crit}}$ & $\beta$ \\ \hline
 $6$ & $2$ & $1.1(2)$ & $0.081(4)$ & $0.33(3)$ \\ 
 $10$ & $2$ & $1.2(1)$ & $0.085(4)$ & $0.34(3)$ \\ 
 $10$ & $3$ & $1.1(1)$ & $0.082(4)$ & $0.31(3)$ 

\end{tabular}
\end{center}
\caption{Fit parameters for vacuum expectation value calculations.}
\label{tabone}
\end{table}
\renewcommand{\arraystretch}{3}

\renewcommand{\arraystretch}{1}
\begin{table}[tbh]
\begin{center}
\begin{tabular}{ccccc}

 $J_{\subtext{max}}$ & $R$ & $b$ & $c_{\subtext{crit}}$ & $\nu$ \\ \hline
 $6$ & $2$ & $3.2(4)$ & $0.064(6)$ & $0.72(7)$ \\ 
 $10$ & $2$ & $3.8(7)$ & $0.077(7)$ & $0.69(9)$ \\ 
 $10$ & $3$ & $3.7(6)$ & $0.074(7)$ & $0.78(9)$ 

\end{tabular}
\end{center}
\caption{Fit parameters for physical mass calculations.}
\label{tabtwo}
\end{table}
\renewcommand{\arraystretch}{3}

\renewcommand{\arraystretch}{1}
\begin{table}[thb]
\begin{center}
\begin{tabular}{ccccc}

 Ref. & Method & $\beta$ & $\nu$ \\ \hline
 \cite{BFMS} & MC & $0.3265(3)(1)$ & $0.6294(5)(5)$ \\ 
 \cite{CPRV} & HT & $0.32648(18)$ & $0.63002(23)$ \\ 
 \cite{GZJ} & $d=3$ exp.\  & $0.3258(14)$ & $0.6304(13)$ \\ 
 \cite{GZJ} & $\epsilon$-exp.\  & $0.3265(15)$ & $0.6305(25)$ 

\end{tabular}
\end{center}
\caption{Recent results for the critical exponents of the 
three-dimensional Ising universality class.}
\label{tabthree}
\end{table}
\renewcommand{\arraystretch}{3}

\section{Summary}

We have used the formalism of spherical field theory to analyze the phase structure of
three-dimensional $\phi^4$ theory.  This analysis is intended as a demonstration 
that this new nonperturbative method can be successfully extended beyond two-dimensional
theories.  We find that for $\mu^2>0$ there is no
spontaneous breaking of the $\phi\rightarrow -\phi$ reflection symmetry while
for $\mu^2<0$ this symmetry is broken at weak coupling and is restored as the
coupling strength is increased.  We have calculated the critical coupling and the
critical exponents $\beta$ and $\nu$ which characterize the phase transition 
in the latter case by analyzing the behavior of the vacuum expectation value 
and the physical mass as functions
of the coupling.  Our result for $\beta$, which was obtained from the vacuum 
expectation value analysis, is in agreement with recent calculations within the 
three-dimensional Ising universality class.
However, our result for $\nu$ is somewhat larger than the results of these calculations. 
This is most likely due to finite-size errors which affect the physical mass
analysis.  

Several methods of improving this analysis come to mind.  Our computations were
performed with limited computing power: a 500 MHz Alpha workstation and a 
350 MHz 
PC processor.  Greater computing power could improve the analysis in two ways.
First, the number of iterations performed within the Monte Carlo algorithm could be
increased thereby reducing the statistical errors in the computations and 
increasing the precision of the results.  Second, since the computation time 
scales roughly as $J_{\subtext{max}}^4$ a substantial increase in computing power
would allow us to take larger values for $J_{\subtext{max}}$ and $R$ in order to 
reduce the systematic errors which arise due to these parameters being finite.  
Even with the limited computing power mentioned above, however, an improved 
analysis can be achieved with
the method of periodic field theory \cite{MRL} which uses a periodic-box
mode expansion in place of the partial wave decomposition of spherical field theory
and has the advantage of a time-independent hamiltonian.    
Improved calculations using this last approach are currently under way. 

\section*{Acknowledgements}

The author is grateful to Eugene Golowich and Dean Lee for many useful conversations 
and guidance.  Support provided in part by the National Science Foundation.

\end{document}